\documentclass[12pt,preprint]{aastex}

\slugcomment{MS \#71854 v. 3: Astrophysical Journal, 2007, in press}

\shorttitle{Monitoring SN~1978K in NGC~1313}
\shortauthors{Smith et al.}

\begin{document}


\title{Multiwavelength Monitoring of the Unusual Ultraluminous
Supernova SN~1978K in NGC~1313 and the Search for an Associated 
Gamma-Ray Burst}


\author{I. A. Smith}
\affil{Department of Physics and Astronomy, Rice University,\\
6100 South Main MS-108, Houston, TX 77251--1892}
\email{iansmith@rice.edu}

\author{S. D. Ryder}
\affil{Anglo-Australian Observatory, PO Box 296, Epping, NSW 1710, Australia}

\author{M. B\"ottcher}
\affil{Astrophysical Institute, Department of Physics and Astronomy,\\
Clippinger Hall 251B, Ohio University, Athens, OH 45701--2979}

\author{S. J. Tingay}
\affil{Centre for Astrophysics and Supercomputing,\\
Swinburne University of Technology, Hawthorn, Victoria, Australia}

\author{A. Stacy}
\affil{University of Texas, Department of Astronomy,\\
1 University Station, C1400, Austin, Texas 78712--0259}

\author{M. Pakull}
\affil{Observatoire de Strasbourg, 11, rue de l'Universit\'e, 
Strasbourg F-67000, France}

\and

\author{E. P. Liang}
\affil{Department of Physics and Astronomy, Rice University,\\
6100 South Main MS-108, Houston, TX 77251--1892}



\begin{abstract}
We discuss our radio (Australia Telescope Compact Array
and Australian Long Baseline Array)
and X-ray ({\it XMM-Newton}) monitoring observations of the unusual
ultraluminous supernova SN~1978K in NGC~1313 at $\sim 25$ years after 
the explosion.
SN~1978K is a rare example of a Type IIn supernova that has remained
bright enough to have long-term X-ray and radio observations.
The observations probe the dense medium that was ejected by the 
progenitor star prior to its explosion; the star might have been 
a Luminous Blue Variable.
The radio imaging shows that the source remains compact, but 
it may be marginally resolved.
The radio monitoring shows deviations from a smooth decay suggesting 
that inhomogeneities are present in the radio emitting region.
It appears that a major change occurred in the mass-loss rate of
the progenitor star shortly before the supernova event.
The X-ray emission between 2000 and 2006 is consistent with the radiation 
coming from two shocks.
All the X-ray data can be fit using the same model (with no systematic 
evolution or short-term variability) but this has a surprising requirement: 
the X-ray emitting regions have a very large abundance of helium.
This would be consistent with the X-ray emitting shocks being located
in a helium-rich layer that was ejected by the progenitor star, or
helium-rich material was ejected in the supernova explosion.
The unusual properties of the supernova motivated a search for
an associated gamma-ray burst (GRB).
We show that SN~1978K was inside the $\sim 4 \sigma$ error box of
GRB~771029.
If this association is correct, the GRB was exceptionally underluminous.
However, the quality of the gamma-ray burst locations at that time was 
poor, and this is likely just a chance alignment.
\end{abstract}


\keywords{
supernovae: individual (SN~1978K) ---
supernova remnants ---
galaxies: individual (NGC~1313) ---
gamma rays: bursts
}

\section{Introduction} \label{intro}

The nearby late-type barred spiral galaxy \objectname{NGC 1313}
has long been a subject of detailed multiwavelength observations.
It appears to be an isolated galaxy at a distance of only 4.13 Mpc
\citep{men02}.
The disk is inclined at 48\arcdeg\ to the line of sight \citep{ryd95}
permitting an excellent view of the whole galaxy.
It has a patchy optical morphology, contains several \ion{H}{1}
supershells \citep{ryd95}, and is rich in young clusters 
\citep{lar99}.
This indicates that it has undergone vigorous irregular star formation.

The diffuse X-ray emission in NGC~1313 is low which has allowed
detailed studies of its point sources.
Observations have been made by {\it Einstein}, {\it ROSAT}, 
{\it ASCA}, {\it XMM-Newton}, {\it Chandra}, and {\it Suzaku} 
\citep[e.g.][]{fab87,ryd93,sto95,col95,col99,sch96,sch99,sch00,sch04,mil98,mil03,mil04,zam04,miz07,muc07,fen07}.

NGC~1313 contains three ultraluminous X-ray sources (ULXs) that are 
offset from the center of the galaxy. 
Two of these -- with X-ray luminosities $\gtrsim 10^{40}$~erg/s --
have been associated with optical ionized nebulae
\citep{pak02,pak03,pak06,liu07}.
Their X-ray spectral and variability properties (with significant 
changes on time scales as short as 1 day) indicate they contain 
accreting black holes. 
Our multiwavelength observations of these two ULXs and the other less
luminous persistent and transient X-ray sources in NGC~1313 will be 
discussed elsewhere \citep{smi07}.

The third ULX in NGC~1313 is the unusual luminous supernova 
\objectname{SN 1978K}.
This was the second supernova to be detected and recognized as a 
supernova from its radio emission and the first from its X-rays
\citep{ryd93}.
There was no X-ray detection by {\it Einstein} in 1980;
the first X-ray detection was by {\it ROSAT}, 13 years after 
the explosion.
It is rare that supernovae are bright enough to be followed in 
the X-ray band \citep{bre03}, and SN~1978K is one of only a few 
that have had long-term monitoring.

There was no evidence for radio emission prior to the supernova.
Unlike a ``normal'' Type II supernova, the peak radio luminosity 
was very high.
SN~1978K is thus one of the most important members of the ``Type IIn'' 
sub-class of supernovae \citep{sch00}.

Observations before the supernova showed the progenitor star
to have $B_J \sim 22.1$ in 1974 and 1976, and it was at 
$B_J \gtrsim 23$ on 1977 October 12, the last observation
prior to the explosion \citep{ryd93}.
The object brightened to $B=16$ by 1978 July 31, and faded to 
$B=18$ by the end of 1978 September \citep{ryd93}.

The optical emission lines in SN~1978K are only moderately broad, 
unlike most supernova ejecta \citep{ryd93,chu95,chu99}.
This suggests the surrounding medium was dense, leading to a rapid 
slowdown of the shock.
It is likely that the dense medium is the wind ejected prior to the
explosion by the progenitor star, which might have been a Luminous 
Blue Variable \citep[LBV; ][]{chu99,gru02}.

{\it Hubble Space Telescope} observations have shown that the source
is still consistent with being point-like \citep{gru02}.
The upper limit of $0.1\arcsec$ on the size of the nebula that
is currently radiating in the optical ($\sim 2~{\rm pc}$) is again 
consistent with the progenitor being an LBV.

In addition to the moderately broad lines, there are narrow optical 
emission lines \citep{chu95,chu99,gru02}.
Some of these likely come from shocked gas in the dense circumstellar 
wind.
\citet{chu95} note the possible presence of the narrow forbidden
lines [\ion{Fe}{10}] and [\ion{Fe}{14}] that might originate from
hot ($\sim 10^6~{\rm K}$) shocked gas.
Other narrow optical emission lines come from a circumstellar 
\ion{H}{2} region that has not yet been reached by the supernova 
ejecta and that might result in a re-brightening in the future.

The evolution of the broad-band radio light curves indicated
there was a time-independent free-free absorption
component required \citep{mon97,sch99}.
This is presumably the same \ion{H}{2} circumstellar region inferred from 
the narrow optical lines.
This provides further evidence for the suggestion that the progenitor
was an LBV.

In this paper, we discuss our radio observations using the
Australia Telescope Compact Array (ATCA) and Long Baseline Array 
(LBA), and X-ray observations using {\it XMM-Newton}
of SN~1978K at $\sim 25$ years after the explosion.
Preliminary results were presented in \citet{smi04} and \citet{sta05}.

In \S\ref{radio} we show the new radio observations of SN~1978K and
discuss the implications for the radio imaging and variability.
In \S\ref{xray} we summarize the X-ray monitoring observations from 
2000 through 2006 and discuss the modeling of this data.
In \S\ref{grb} we describe a search for a gamma-ray burst (GRB) that 
might have accompanied the supernova.
In \S\ref{conclude} we conclude with the implications for future 
observations.

\section{Radio Observations} \label{radio}

\subsection{VLBI Imaging Observations} \label{vlbi}

SN~1978K was observed with an Australian VLBI array consisting of the
following antennas: Tidbinbilla 70~m (NASA); ATCA $5 \times 22$~m
phased array (ATNF); Parkes 64~m (ATNF); Mopra 22~m (ATNF); and Hobart
26~m (University of Tasmania).

A 12 hour observation using all the antennas except Tidbinbilla
took place on 2003 February 14 between 00:00 UT and 12:00 UT.
This used two 16~MHz bands in the frequency ranges $1.384 - 1.400$~GHz 
and $1.400 - 1.416$~GHz.
A second 12 hour observation using all the antennas took place on 
2003 March 22, also between 00:00 UT and 12:00 UT.
This used two 16~MHz bands in the frequency ranges $2.264 - 2.280$~GHz 
and $2.280 - 2.296$~GHz.

For both observations the Nyquist sampled, two-bit digitized right
circular polarization signals were recorded at each antenna using the
S2 tape-based system \citep{can97}.
All the data were correlated at the ATNF S2 correlator \citep{wil92},
and the correlated data were exported to {\sc aips} for standard 
fringe-fitting and amplitude calibration using measured $T_{\rm sys}$ 
values from the telescopes.  
These data were exported to {\sc difmap} \citep{she94} for further 
calibration and imaging, using standard self-calibration and clean 
techniques.

\begin{figure}
\epsscale{.80}
\includegraphics*[angle=270,bb=65 172 521 623]{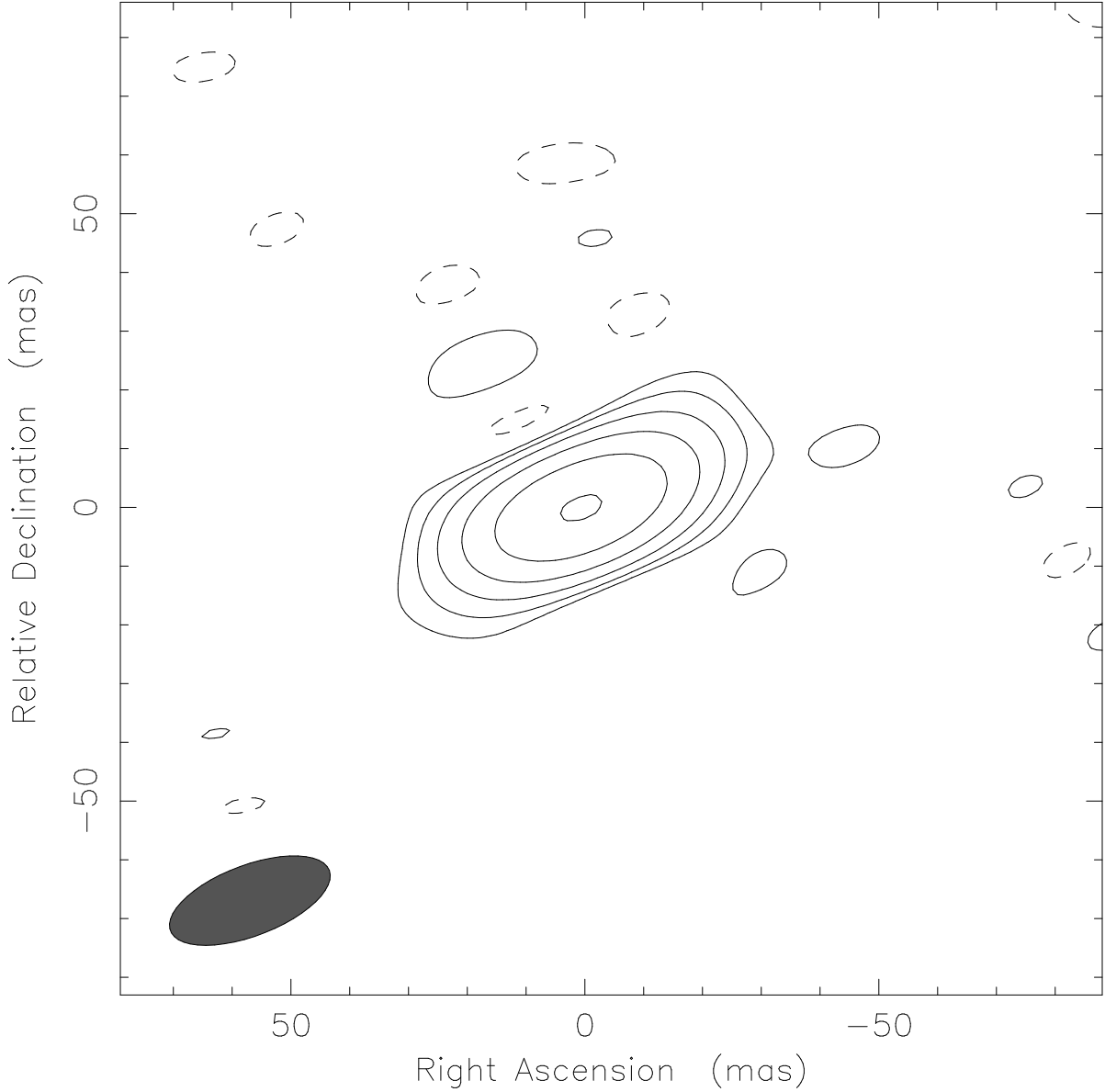}
\caption{VLBI image of SN~1978K at a frequency of 2.3 GHz 
on 2003 March 22. 
The map is centered at 03:17:38.62, $-66$:33:03.4 (J2000).
The peak flux density in the image is 28 mJy/beam and the
rms image noise is 0.5 mJy/beam.  
The beam is 29 mas $\times$ 12 mas at a position angle of $-69\arcdeg$.  
Contours are shown at $\pm3$\%, 6\%, 12\%, 24\%, 48\%, and 96\% of the 
peak flux density.
\label{vlbifigure}
}
\end{figure}

Figure \ref{vlbifigure} shows the final image from the higher 
resolution 2.3~GHz data.  
A single compact component was detected at a signal to noise of 
$\sim 56$ (the peak flux density was $28 \pm 3$~mJy/beam, and the 
rms image noise 0.5~mJy/beam). 
The total cleaned flux in this component is $33 \pm 3$~mJy.  
The component appears to be marginally resolved.  
When modeled as a single circular Gaussian component in the (u,v) 
plane, the best-fit FWHM is $9.5 \pm 2.0$~mas (c.f. beam dimensions 
of 29~mas~$\times$~12~mas).  

At 1.4~GHz the source appears very similar, being marginally resolved
and having a best-fit circular Gaussian FWHM of $12 \pm 3$~mas (c.f.
beam dimensions of 38~mas~$\times$~17~mas).  
The estimated sizes at 1.4~GHz and 2.3~GHz are consistent to within 
their errors.  
The total flux density of the marginally resolved component 
at 1.4~GHz is higher than at 2.3~GHz, $56 \pm 6$~mJy.
The source is detected at a signal to noise of $\sim 45$ (1.4~GHz 
image rms of 1.0~mJy/beam and peak flux density in the image of 
$45 \pm 5$~mJy/beam).

\subsection{ATCA Monitoring Observations} \label{atca}

Following on from the previous radio observations 
\citep{ryd93,mon97,sch99}, the ATCA has continued to make
occasional observations of SN~1978K.
The data acquisition followed the same procedures as outlined 
in \citet{sch99}.
The dataset for each observation and frequency was edited and
calibrated using the {\sc miriad} software package. 
Rather than estimating the unresolved flux density of SN~1978K from 
plots of the calibrated visibility amplitude as in \citet{sch99}, 
the {\sc uvfit} task was used instead in the visibility domain to 
fit a point source at the known location of SN~1978K.

\begin{figure}
\plotone{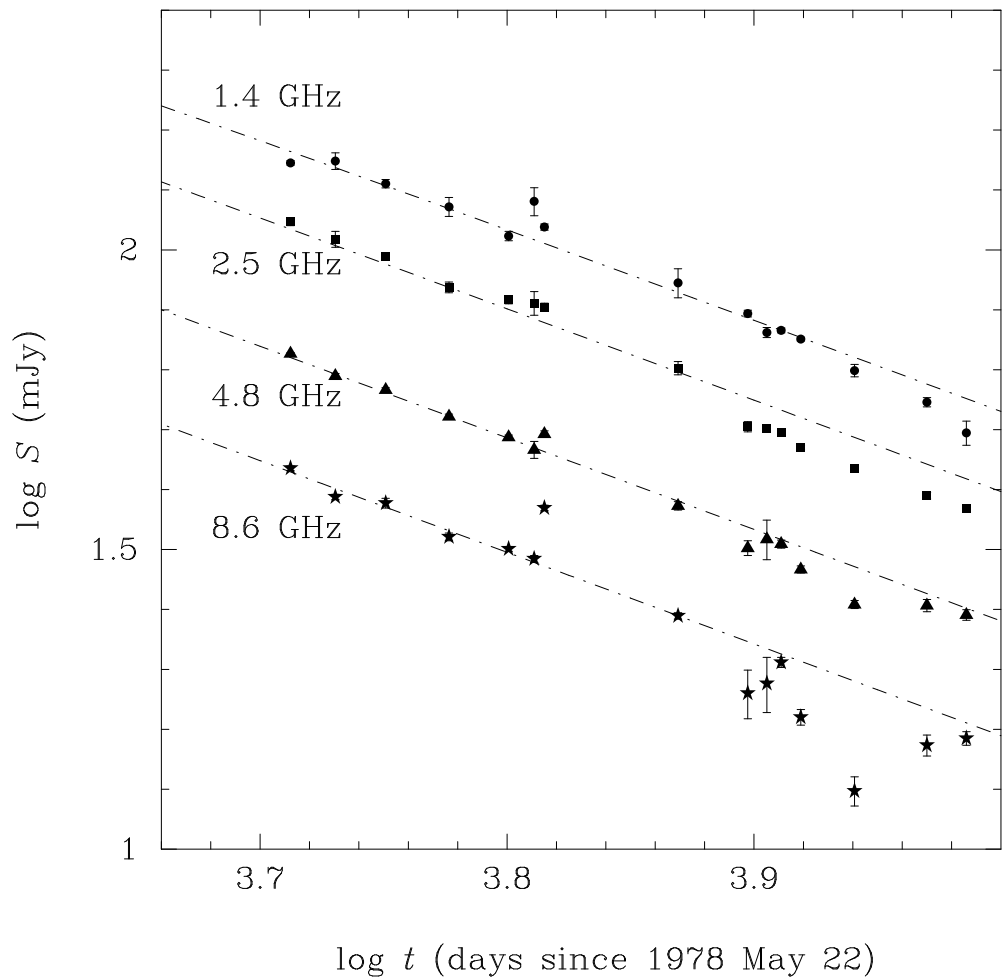}
\caption{The radio light curves for SN~1978K from ATCA monitoring 
between 1991 and 2004.
This adds seven points to each of the curves shown in \citet{sch99}.
The curves are the best fits to the first 8 epochs of radio 
observation using a modified version of the \citet{wei90} model 
for the evolution of Type II supernovae described in \citet{mon97}.
\label{atcafigure}
}
\end{figure}

Figure \ref{atcafigure} and Table \ref{radiotable} add 7 
data points at all four frequencies (8640, 4790, 2496, and 1376 MHz)
to the results presented by \citet{sch99}.
The ``age'' assumes that the supernova had a maximum on
1978 May 22 (MJD 43650).

The curves in Figure \ref{atcafigure} are the best fits to 
the first 8 epochs of radio observation \citep{sch99}
using a modified version of the \citet{wei90} model for the 
evolution of Type II supernovae described in \citet{mon97}.
Figure \ref{atcafigure} shows that while the evolution has generally 
continued to follow the model curves, there have been significant 
deviations from a smooth decay.
In particular, there was a major dip at higher frequencies in April 2002.
The most recent observation suggests a return to the long term
rate of decline at each frequency.

\subsection {Radio Implications} \label{radiovar}

The most recent radio observations of SN~1978K show that it continues 
to be easily detectable more than a quarter of a century after the
explosion, but with increasing deviations from a smooth decline. 

The first tentative VLBI detection of a resolved remnant
$\sim 10$~mas in size would correspond to a diameter of 0.2~pc, 
which allows us to place an upper limit on the average expansion 
velocity of $4000~{\rm km~s}^{-1}$.

The significant departures from a smooth decline likely reflect 
changes in the circumstellar medium density.
This indicates that a major change occurred in the mass-loss rate 
of the progenitor star prior to the supernova event.
The time that this change took place depends on the assumed stellar 
wind velocity, and thus the progenitor type. 
For a Wolf-Rayet/LBV progenitor with a stellar wind 
$\sim 2000~{\rm km~s}^{-1}$, the change in mass-loss rate must 
have occurred $\sim 4000/2000 \times 25 = 50$ years before
the explosion. 
For a red supergiant wind speed of $\sim 20~{\rm km~s}^{-1}$
the change would have occurred 5,000 years earlier.

\citet{chu95} suggested that variations in the ${\rm H} \alpha$
luminosity were similarly due to a clumpy circumstellar material.
Assuming a wind speed of $\sim 10~{\rm km~s}^{-1}$, they inferred
that this could be explained by changes in the mass-loss 
$3,000 - 10,000$~years prior to the explosion.

\section{X-ray Observations} \label{xray}

NGC~1313 is close enough that {\it XMM-Newton} can simultaneously 
perform detailed spectroscopic and timing studies of all the ULXs in 
short observations. 

\subsection{{\it XMM-Newton} Observations} \label{xmm}

In {\it XMM-Newton} Cycles 2 and 3, we performed a series of 
$\sim 10~{\rm ksec}$ Priority A observations to investigate the 
evolution of the ULXs and the other X-ray sources in NGC~1313
on a variety of time scales.

In Cycle 2, eight observations took place between 2003-11-25
and 2004-01-16.
Unfortunately, several of these were affected by soft 
proton flaring that limit their usefulness for detailed analysis.
In Cycle 3, we observed approximately every three months 
(the first observation was redone due to a slew failure) 
to monitor the longer-term variability.

All the Cycle 2 and 3 observations were centered on ULX X-2.
An advantage of using {\it XMM-Newton} to observe NGC~1313 is 
that ULX X-1, X-2, and SN~1978K are all well within the field 
of view in a single pointing.
However, depending on the satellite orientation, in some of the 
observations SN~1978K was located on or very close to a bad column
in one of the detectors, or on or near a boundary between MOS 
or pn CCDs: in these cases, little or no useful data was collected 
by one or more of the detectors.

The {\it XMM-Newton} observation on 2003-12-09 was fully 
simultaneous with a portion of the ATCA observation.
Unfortunately, this {\it XMM-Newton} observation was plagued
by flares, and none of the data taken that day are useful.
The {\it XMM-Newton} observation on 2004-11-23 was fully 
simultaneous with a portion of the ATCA observation.
The whole of this {\it XMM-Newton} observation was unaffected by flares.
However, a dark column ran through SN~1978K in the pn data, and thus
only the MOS1 and MOS2 data are useful for that day.

In Table \ref{xraytable} we list all our {\it XMM-Newton} observations
from Cycles 2 and 3.
We have also re-analyzed the observation of NGC~1313 from 2000-10-17 
(PI: Bernd Aschenbach).
For each observation, we list the detectors that were used in the 
modeling of SN~1978K in \S \ref{xmodel}, and the cleaned exposure times.
While some of these spectra have short exposure times due to the
flaring, leaving out the poorest spectra does not make any significant 
difference to the fitting results.

All the 2003-2005 observations were made in Full Frame EPIC mode, and used
the thin optical blocking filter.
Although the sources in NGC~1313 are bright, pile-up was not a problem 
for any of them.
The total count rate limits for the detectors were not reached 
thanks to the low diffuse emission from the galaxy.

All the 2000-2005 data were reduced using version 6.5 of the {\it XMM-Newton}
Science Analysis System (SAS).
A couple of test cases were re-reduced using the newer version 7.0
of SAS, but this did not result in any significant changes.
The spectral and timing analysis was performed using HEAsoft 
version 6.1.

The data reduction followed the standard steps in SAS.
For all the data, the pipeline processing was redone using emchain
and epchain.
The times were determined when soft proton flaring was significant
-- when the full-image count rate for photons $> 10~{\rm keV}$ was 
above 0.35~cps for the MOS and above 1~cps for the pn -- and these 
data segments were removed.
Images were made for each detector, and it was examined whether
SN~1978K was significantly affected by bad columns on the CCD or 
by the boundary between CCDs.
There were no Out of Time problems for the source.
Spectra and light curves were made for SN~1978K using the 0-12 X-ray 
patterns for the MOS and the 0-4 patterns for the pn.
A $40\arcsec$ extraction radius around the source was used.
The data were screened for defects with \#XMMEA\_EM (MOS) or 
\#XMMEA\_EP (pn), and events next to the CCD edges were excluded
using FLAG=0.
Background spectra and light curves were made in the same fashion
using a nearby region on the same CCD that was free of defects.
Since the telescope orientation was not the same for each observation,
the location of the background region changed with time;
however, this was not a significant problem since the background 
flux was always much lower than that of the source.
Photon redistribution matrices (RMF) and ancillary region files (ARF)
were generated for the location of SN~1978K for use in the spectral 
analysis.
The calibration of the EPIC instruments was good enough that 
no fixed or floating normalization factors were needed to match 
the fluxes for the three detectors; the improved cross-calibration
of the detectors in SAS 6.5 fixed the problems in our initial 
analysis \citep{sta05}.

After our original manuscript was submitted, an additional 
{\it XMM-Newton} observation of NGC 1313 taken on 2006-03-06 
became publicly available (PI: Stefan Immler).
This observation was long and clean -- with very little soft proton 
flaring -- and was centered on SN 1978K.
The medium optical blocking filter was used in this observation.
The 2006-03-06 data were reduced with SAS 7.0 in the same manner as 
the other data.
We found that the 2006-03-06 observation is well fit by the model 
described in \S \ref{xmodel} that was used to fit the 2000-2005 data.
Folding the 2006 data into the joint 2000-2005 fit leads to very 
small changes in the ``best fit'' values cited in \S \ref{xmodel};
the optimized values are all well within the original errors for the 
parameters.
Rather than re-writing all of \S \ref{xmodel}, we have left that
section as a discussion of the modeling of the 2000-2005 data.
We have added \S \ref{x2006} to discuss the fitting of the 2006 data,
which acts as an independent test of the model.

\subsection{Modeling 2000-2005 {\it XMM-Newton} Data} \label{xmodel}

The X-ray emission from SN~1978K has been fit using an absorbed
two-temperature optically thin hot thermal plasma model 
(dual vmekal in XSPEC) to model the forward and reverse shocks 
formed by the interaction of the supernova ejecta with its 
surroundings \citep{sch04}.

Modeling the {\it XMM-Newton} spectra for each of the separate days,
we find that acceptable statistical fits can be obtained using a
range of model parameters.
Since fluctuations in the X-ray light curve with amplitudes $\sim 20$\% 
might be seen if the ejecta are significantly inhomogeneous
(or more based on the radio variability), we do not 
necessarily expect to get the same fit results for each day.
It is also plausible that the X-ray light curve might be decaying 
as a power law as the remnant expands \citep{sch99}, or it might
brighten steadily if the supernova ejecta runs into a denser
circumstellar envelope.
There could also be other systematic variations -- such as the 
temperatures cooling -- in one or more of the parameters.
However, since several of our observations span a relatively short 
range of time compared to the time since the explosion, significant 
changes in the model parameters may be less likely.
Historically, the recent X-ray monitoring of SN~1978K has found it
consistent with being constant \citep{sch00}.

To address the variability issue in a concrete way, we have combined
all the {\it XMM-Newton} data from 2000 to 2005 and investigated 
whether there is a model that can simultaneously fit all this data.
This assumes that there is no systematic evolution or significant
short-term variability for any of the fit parameters.
If we were unable to find a good fit to the joint data, this would 
imply that one or more of the parameters was changing significantly.
On the other hand, finding a good fit to the joint data does not
necessarily guarantee that this is the correct result; the spectra 
from a variable source might by chance average out (although the 
increase in the number of photons in the joint fit makes this 
difficult to accomplish).
Ultimately, as we illustrate in \S \ref{x2006}, the joint fit can 
be fully tested by using it to fit future observations, allowing 
for plausible slow long-term evolutions.

\begin{figure}
\plotone{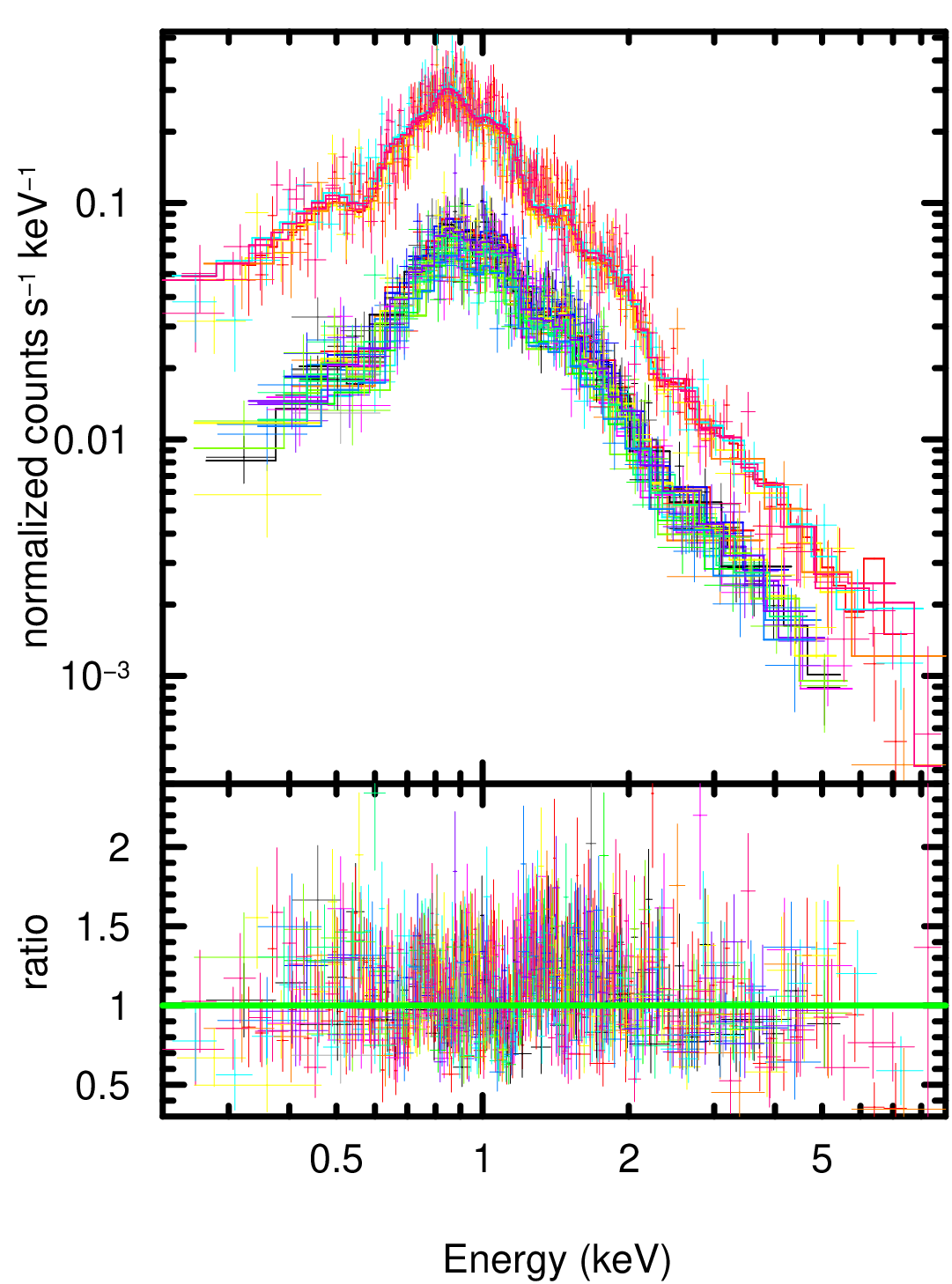}
\caption{Fit to all the {\it XMM-Newton} observations of SN~1978K 
from 2000 through 2005 using an absorbed dual vmekal model with 
solar abundances in the hot plasmas.
See the electronic edition of the Journal for a color version of this figure.
\label{xmmsolarfigure}
}
\end{figure}

Our modeling of the combined data found that it was not possible to get 
a good fit using solar abundances for the hot plasmas.
Assuming solar abundances for all the components, the best fit gives
a reduced chi squared of $\chi^{2}_{\nu} = 1.26$
for $\nu = 1139$ degrees of freedom and an unacceptably low null 
hypothesis probability of $Q = 7.6 \times 10^{-9}$.
This is shown in Figure \ref{xmmsolarfigure}.
Although the temperatures of the vmekal components are similar
to previous results \citep{sch04}, the value of 
$N_{\rm H} = 1.16 \times 10^{21}~{\rm cm}^{-2}$ in this case
is rather low.

\begin{figure}
\plotone{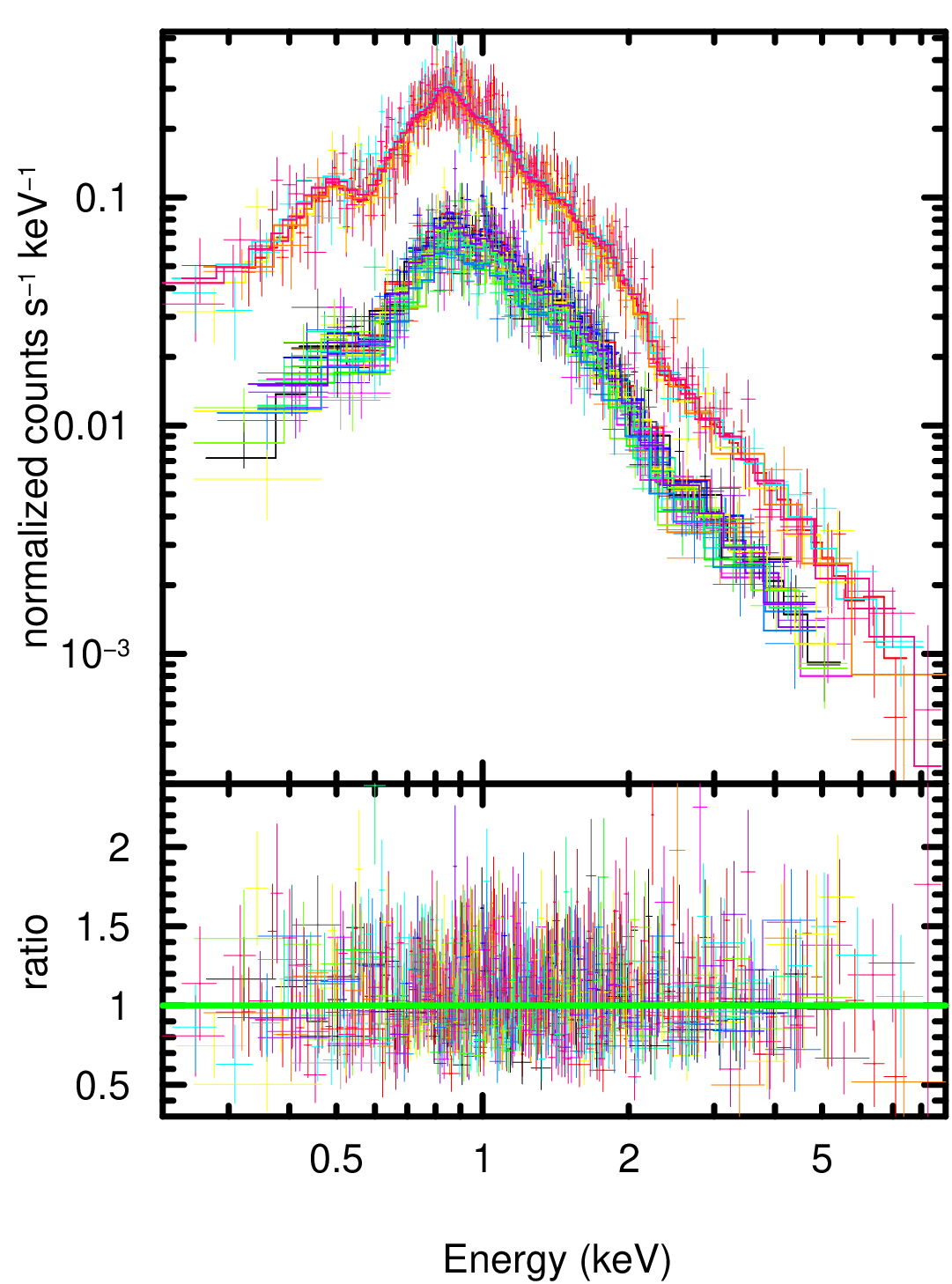}
\caption{Fit to all the {\it XMM-Newton} observations of SN~1978K 
from 2000 through 2005 using an absorbed dual vmekal model with 
large helium abundances in the hot plasmas.
See the electronic edition of the Journal for a color version of this figure.
\label{xmmfitfigure}
}
\end{figure}

We investigated varying the abundances of the vmekal model components.
We systematically tried allowing individual elements to vary without 
constraint (with the absorption, temperatures, and normalizations
also unconstrained).
We also tried allowing combinations of elements to vary without 
constraint.
Surprisingly, we found that acceptable fits could only be attained in
one special circumstance: the helium abundances in both the plasmas 
need to be large.
This would imply that the X-ray emitting shocks are currently located
in a helium-rich layer that was ejected by the progenitor star, 
or helium-rich material was ejected in the supernova explosion.
A representative fit is shown in Figure \ref{xmmfitfigure}.
The parameters for this fit are given in Table \ref{jointmodel}.
With the exception of the helium abundances, the fit parameters
are similar to those found previously by \citet{sch04}.
The fit has $\chi^{2}_{\nu} = 1.06$ for $\nu = 1137$ and $Q = 0.084$.
The ratio of the model to the data does not have systematic
variations, unlike the case when using solar abundances.

Fixing all the parameters of the large helium abundance model to the 
values given in Table \ref{jointmodel}, we compared the model to 
the data for each day separately to see if every day could be 
explained by the same model.
Only one day gave an unacceptable result; 2004-05-01 had 
$Q = 1.8 \times 10^{-4}$.
However, only MOS data was available for that day.
For 2004-05-01, leaving all the parameters fixed at the values in 
Table \ref{jointmodel} but allowing just the helium abundances to 
vary gave a good fit ($Q = 0.24$) with helium abundances of 7.5 
for the hard component and 23.5 for the soft;
these values are consistent with the original error bars for these 
parameters.
The following observation on 2004-06-05 -- that had more data -- was 
well explained by the original fixed model ($Q = 0.44$).
Thus if there was any deviation in the abundances, it was short-lived.
The next worst day 2003-12-21 had $Q = 0.16$, which is a fine fit.
Thus the individual day results are consistent with the assumption 
that there was no systematic evolution or large-scale variability.

For SN~1978K, the absorption found in our best joint fit is
$N_{\rm H} = (2.12^{+0.14}_{-0.16}) \times 10^{21}~{\rm cm}^{-2}$.
This is much higher than the value typically found for the rest of 
NGC~1313 plus the Milky Way 
($N_{\rm H} = 0.37\times 10^{21}~{\rm cm}^{-2}$).
Thus this must be dominated by absorbing material close to SN~1978K.
The value of $N_{\rm H}$ found in our joint fit is tightly 
constrained and agrees very well with the value of 
$N_{\rm H} = (2.2 \pm 0.1) \times 10^{21}~{\rm cm}^{-2}$
that was found from the mean \ion{H}{1} column density 
measured around SN~1978K \citep{ryd93}.
This adds further credence to our joint fit results and the large helium
abundance model.
The implication of our model is that the absorption is currently 
dominated by the material ejected by the supernova progenitor and/or 
its natal molecular cloud.

\citet{sch04} had a range of $N_{\rm H}$ in their different fits and 
suggested that $N_{\rm H}$ may be increasing with time.
We instead conclude that $N_{\rm H}$ is consistent with remaining 
constant over the recent observations.

\citet{sch04} also suggested that an excess of silicon improved the fit.
However, this only affects a couple of spectral bins and we do not
find any significant improvements in the fit when the silicon
abundances are varied.

We have studied varying the metallicity of the absorbing material
for SN~1978K, but find no significant deviations from solar abundances.
The fact that the absorption is currently dominated by material 
close to the source means that the metallicity of the interstellar
medium in NGC~1313 is not an important issue when fitting the 
SN~1978K X-ray spectra, unlike the case for the accreting ULXs.
A low metallicity absorption is indicated when fitting the 
{\it XMM-Newton} data for the two accreting ULXs 
\citep{smi04,smi07,miz07}.
This is consistent with other studies that have shown that the metallicity 
in NGC~1313 is a factor of 4 lower than in the Milky Way 
and that this is the highest mass barred spiral without a radial 
abundance gradient \citep{wal97,mol99}.
Allowing for a low abundance of oxygen is particularly important
when modeling the low temperature soft component in ULXs which has been 
interpreted as being disk blackbody emission from an optically thick 
disk around a $\sim 1000 \, M_{\odot}$ black hole.

The unabsorbed fluxes from SN~1978K in the $0.5-2~{\rm keV}$ 
and $2-10~{\rm keV}$ bands are 
$7.6 \times 10^{-13}~\rm{ergs}~{\rm cm}^{-2}~{\rm s}^{-1}$
and $3.2 \times 10^{-13}~\rm{ergs}~{\rm cm}^{-2}~{\rm s}^{-1}$
respectively.
These fluxes are similar to those in \citet{sch04}.
Assuming a distance of 4.13 Mpc \citep{men02}, the corresponding 
luminosities are $1.6 \times 10^{39}~\rm{ergs}~{\rm s}^{-1}$
($0.5-2~{\rm keV}$) and $6.5 \times 10^{38}~\rm{ergs}~{\rm s}^{-1}$
($2-10~{\rm keV}$).
The luminosity for the whole $0.2-10~{\rm keV}$ band is
$2.9 \times 10^{39}~\rm{ergs}~{\rm s}^{-1}$.

\subsection{Testing the Model using the 2006 {\it XMM-Newton} 
Data} \label{x2006}

The long clean 2006-03-06 {\it XMM-Newton} observation of SN 1978K was 
used as an independent test of the results found in \S \ref{xmodel}.

As was found when fitting the 2000-2005 data, the 2006-03-06 data
cannot be fit using a dual vmekal model with solar abundances for 
the hot plasmas.
The best fit in this case gives $\chi^{2}_{\nu} = 1.42$ for 
$\nu = 283$ and an unacceptably low $Q = 4.7 \times 10^{-6}$.

The 2006-03-06 observation is well fit by the large helium abundance
model described in \S \ref{xmodel} that fit the 2000-2005 data.
Using the parameters from Table \ref{jointmodel} -- with no additional
optimization -- the fit has $\chi^{2}_{\nu} = 1.09$ for $\nu = 281$ 
and $Q = 0.14$.

Folding the 2006 data into the joint 2000-2005 fit leads to very 
small changes in the ``best fit'' values for the large helium
abundance model given in Table \ref{jointmodel}.
The new values are all well within the original errors for the 
parameters.

These results bolster the conclusions that (1) there has been no 
significant evolution of the X-ray parameters recently, and (2) the 
large helium abundance model gives a good explanation for the 
2000-2006 data.

\section{Search for an Associated Gamma-Ray Burst} \label{grb}

In models for long-duration GRBs, it is thought that the 
progenitor star at least ejects its hydrogen envelope 
prior to the supernova/hypernova \citep[e.g.][]{woo06}. 
If this did not take place, the relativistic jet would not escape 
with sufficient energy to make a long GRB.
However, only a tiny fraction of massive stars produce a GRB when 
they die, perhaps $\sim 1 \%$ \citep{gue07}.
This leads to the question of what features of the progenitor
are necessary/sufficient to produce a GRB.
Although GRBs are believed to be associated with Type Ib or Ic
supernovae, given the rare and unusual nature of the Type IIn 
SN~1978K and the fact that it had significant ejections prior
to the explosion, we have searched for any GRBs that might have 
been associated with this supernova.

Based on the optical observations summarized in \S \ref{intro}, 
the supernova explosion took place some time between 1977 October 12 
and 1978 July 31.
These dates bracket our search for an associated GRB.
The Interplanetary Network (IPN) of GRB satellites has existed since 1976.
Several GRB detectors were operational during the window for SN~1978K.
However, the first wide-baseline three-cornered network did not exist
until the launch of the Veneras in 1978 September \citep{att87}.
Thus the localization of the bursts was poor during the SN~1978K window.

\citet{kle82} used the data from the available satellites to determine
the GRBs that were detected by at least two instruments on separate 
satellites.
Other events may have been real bursts, but only the confirmed GRBs
were reported.
In the SN~1978K window, 6 confirmed GRBs were detected: GRB~771020,
GRB~771029, GRB~771110, GRB~780508, GRB~780519, and GRB~780521.

Based on the locations of these bursts \citep{kle82}, 5
of them are clearly not consistent with the location of SN~1978K.
However, GRB~771029 is close to SN~1978K; the supernova
is inside the $\sim 4 \sigma$ error box of GRB~771029.
Unfortunately, the uncertainty in the location for this burst 
is large, and the probability that SN~1978K would lie in the 
$4 \sigma$ error annulus is 14\%.
Thus this may be just a chance alignment.

GRB~771029 was a long-duration burst, lasting $\sim 15~{\rm sec}$
\citep{est78}.
It was detected in the $140-300~{\rm keV}$ band by SIGNE 2MP detector S
on Prognoz 6.
GRB~771029 emitted $< 10^{-5}~{\rm erg}~{\rm cm}^{-2}$ and so was 
less energetic than the other two Prognoz 6 events
($> 2 \times 10^{-5}~{\rm erg}~{\rm cm}^{-2}$).
Assuming an isotropic gamma-ray emission and a distance of 4.13 Mpc, 
the burst energy would have been $< 2 \times 10^{46}~{\rm ergs}$.
There are currently 3 examples of highly underluminous long-duration
GRBs associated with relatively nearby supernovae \citep{cob06}.
The least luminous was GRB~980425/SN~1998bw that had an isotropic
gamma-ray energy of $\sim 10^{48}~{\rm ergs}$.
Thus if GRB~771029 was associated with SN~1978K, this would have 
been an exceptionally underluminous burst.
However, we do not know what luminosity should be expected for a 
GRB associated with a Type IIn supernova.

An argument against the association of GRB~771029 and SN~1978K is that 
the radio and optical emissions likely peaked in 1978 May, which
would be many months after the burst. 
The existence of lags between the GRB and supernova events is not 
currently supported by observations. 
For example, in the GRB~060218/SN~2006aj association \citep{cam06},
the supernova and GRB were coeval to within $\lesssim 1$ day.

\section{Conclusions} \label{conclude}

The rare Type IIn supernova SN~1978K remains bright in radio
and X-rays $\sim 25$ years after the explosion.
The $0.2-10~{\rm keV}$ luminosity is
$2.9 \times 10^{39}~\rm{ergs}~{\rm s}^{-1}$.

The radio imaging shows that the source remains compact, but 
it may be marginally resolved, $\sim 10$~mas in size corresponding
to a diameter of 0.2~pc.
The recent radio observations show deviations from a smooth decay 
suggesting that inhomogeneities are present in the radio emitting 
region.
It appears that a major change occurred in the mass-loss rate of
the progenitor star -- possibly a LBV -- shortly before
the supernova event.

The winds from the progenitor star produced a dense medium.
The X-ray emission is consistent with coming from two shocks in
this medium.
The {\it XMM-Newton} observations from 2000 to 2006 can all be fit 
using the same model, with no systematic evolution or short-term 
variability.
While the model parameters that we find are similar to those in
previous studies, the new element of our modeling is the
large abundance of helium.
This suggests that the shocks are located in a helium-rich layer 
that was ejected by the progenitor star, or helium-rich material 
was ejected in the supernova explosion.

Using the large helium abundance model, the value of $N_{\rm H}$ is 
tightly constrained and agrees very well with the value found from 
the mean \ion{H}{1} column density measured around SN~1978K.
This indicates that the absorption is currently dominated by the 
material ejected by the progenitor and/or its natal molecular cloud.
The value of $N_{\rm H}$ is consistent with remaining constant 
over the recent observations.

The unusual properties of the supernova motivated a search for
an associated GRB.
Considering just the confirmed GRBs, only one GRB is consistent with
the location of SN~1978K; the supernova is inside the $\sim 4 \sigma$ 
error box of GRB~771029.
However, this is likely just a chance alignment.

It is of particular interest to continue to study the long-term 
multiwavelength evolution of this rare supernova to study the 
mass-loss history of the progenitor star.
By continuing to watch how SN~1978K evolves, it will be possible
to understand whether the unusually dense circumstellar medium
is the result of an extreme mass-loss rate, a low wind velocity,
or the presence of cavities and clumps.
Is the mass loss at late epochs episodic (resulting in an onion-like
shell structure)?
Is it bipolar, with rings and hotspots as in SN~1987A?
Is there evidence that a GRB took place, even if the burst was not 
beamed towards Earth?
Ultimately, is it the environment or intrinsic factors that
dictates which stars end their lives as Type IIn supernovae?
While it is expected that there will be a slow evolution ($\sim$~years)
of the plasma temperatures, normalizations, and/or abundances, the 
variability of the radio indicates that changes on shorter time scales 
should continue to be investigated.


\acknowledgments

This work has been supported by NASA grants NNG04GC64G and NNG04G100G
at Rice University and NNG04GC65G and NNG04GI50G at Ohio University.

The Australia Telescope Compact Array and Australian Long Baseline 
Array are part of the Australia Telescope which is funded by the 
Commonwealth of Australia for operation as a National Facility managed 
by CSIRO.

{\it XMM-Newton} is an ESA science mission with instruments and 
contributions directly funded by ESA Member States and NASA.

We thank the referee for helpful comments.



{\it Facilities:} 
\facility{ATCA}, 
\facility{Hobart},
\facility{Mopra},
\facility{Parkes},
\facility{Tidbinbilla},
\facility{XMM}.

\clearpage

\begin{deluxetable}{lcccccc}
\tablecaption{ATCA radio observations of SN~1978K
\label{radiotable}}
\tablewidth{0pt}
\tablecolumns{7}
\tablehead{
\colhead{UT Date} & 
\colhead{MJD} & 
\colhead{Age} &
\multicolumn{4}{c}{Flux Densities (mJy)} \\
\colhead{} &
\colhead{} &
\colhead{(days)} &
\colhead{1376 MHz} & 
\colhead{2496 MHz} & 
\colhead{4790 MHz} &
\colhead{8640 MHz}
}
\startdata
2000-01-04 & 51547 & 7897 & $78.3\pm1.0$ & $50.7\pm1.0$ & $31.8\pm0.9$ &
$18.2\pm1.7$ \\
2000-05-25 & 51689 & 8039 & $72.8\pm1.4$ & $50.4\pm0.5$ & $32.9\pm2.5$ &
$18.9\pm2.0$ \\
2000-09-10 & 51797 & 8147 & $73.4\pm0.7$ & $49.6\pm0.4$ & $32.3\pm0.5$ &
$20.5\pm0.4$ \\
2001-02-07 & 51948 & 8298 & $71.0\pm0.5$ & $46.8\pm0.4$ & $29.3\pm0.4$ &
$16.6\pm0.5$ \\
2002-04-09 & 52374 & 8724 & $62.9\pm1.5$ & $43.2\pm0.4$ & $25.6\pm0.4$ &
$12.5\pm0.7$ \\
2003-12-09 & 52983 & 9333 & $55.7\pm1.0$ & $38.9\pm0.5$ & $25.5\pm0.6$ &
$14.9\pm0.6$ \\
2004-11-23 & 53333 & 9683 & $49.5\pm2.3$ & $37.1\pm0.5$ & $24.6\pm0.5$ &
$15.3\pm0.4$ \\
\enddata
\end{deluxetable}

\clearpage

\begin{deluxetable}{llcc}
\tablecaption{{\it XMM-Newton} observations of SN~1978K
\label{xraytable}}
\tablewidth{0pt}
\tablecolumns{4}
\tablehead{
\colhead{Start Date (UT)} & 
\colhead{Observation ID} & 
\colhead{Detectors with good data} &
\colhead{Exposure times (ksec)} \\
}
\startdata
2000-10-17 & 0106860101 &       MOS2, pn &       26.1, 20.2 \\
2003-11-25 & 0150280101 & MOS1, MOS2     &  2.5,  2.5       \\
2003-12-09 & 0150280201 &       none     &                  \\
2003-12-21 & 0150280301 & MOS1, MOS2, pn & 10.2, 10.3,  7.6 \\
2003-12-23 & 0150280401 & MOS1, MOS2     &  6.2,  6.8       \\
2003-12-25 & 0150280501 & MOS1           &  8.4             \\
2003-12-27 & 0150280701 &       none     &                  \\
2004-01-08 & 0150280601 & MOS1, MOS2, pn & 12.7, 13.0,  7.1 \\
2004-01-16 & 0150281101 & MOS1, MOS2     &  7.8,  7.7       \\
2004-05-01 & 0205230201 & MOS1, MOS2     &  7.8,  8.0       \\
2004-06-05 & 0205230301 & MOS1, MOS2, pn & 11.5, 11.5,  8.9 \\
2004-08-23 & 0205230401 & MOS1, MOS2, pn & 13.1, 13.9,  4.4 \\
2004-11-23 & 0205230501 & MOS1, MOS2     & 15.5, 15.5       \\
2005-02-07 & 0205230601 & MOS1, MOS2, pn & 12.2, 12.4,  8.5 \\
2006-03-06 & 0301860101 & MOS1, MOS2, pn & 21.3, 21.3, 17.5 \\
\enddata
\end{deluxetable}

\clearpage

\begin{deluxetable}{llc}
\tablecaption{Parameters used in the absorbed dual vmekal modeling of 
the joint fit to the 2000-2005 {\it XMM-Newton} observations of SN~1978K.
All the other abundance parameters are set to 1.0.
Each error gives an estimate of the 90\% confidence region for 
a single interesting parameter.
\label{jointmodel}}
\tablewidth{0pt}
\tablecolumns{3}
\tablehead{
\colhead{Model component} & 
\colhead{Parameter} & 
\colhead{Best fit}\\
}
\startdata
vphabs      & $N_{\rm H}$ ($\times 10^{21}~{\rm cm}^{-2}$) & $2.12^{+0.14}_{-0.16}$ \\
\\
soft vmekal & ${kT}$ (keV)             & $0.61^{+0.02}_{-0.02}$ \\
            & He                       & $39.2^{+9.5}_{-9.2}$ \\
            & Norm ($\times 10^{-4}$)  & $0.92^{+0.06}_{-0.05}$ \\
\\
hard vmekal & ${kT}$ (keV)             & $3.37^{+0.50}_{-0.42}$ \\
            & He                       & $8.9^{+9.2}_{-2.8}$ \\
            & Norm ($\times 10^{-4}$)  & $1.36^{+0.38}_{-0.61}$ \\
\enddata
\end{deluxetable}

\clearpage

\clearpage

\clearpage

\clearpage

\end{document}